\documentclass[pra, reprint, twocolumn, superscriptaddress, amsmath, amssymb, aps, ,floatfix]{revtex4-2}

\usepackage[english]{babel}
\usepackage{graphicx}
\graphicspath{{./Images/},{./ImagesAppendix/},{./images/},{./imagesAppendix/}}

\usepackage{amsmath, amssymb, amsfonts}
\usepackage{physics}
\usepackage{bm}
\usepackage{braket}
\usepackage{array}
\usepackage{multirow}
\usepackage{float}
\usepackage{commath}
\usepackage{verbatim}
\usepackage{color}
\usepackage{enumitem}
\usepackage{array}   
\usepackage{dsfont}
\usepackage{mathtools}
\usepackage{soul}
\usepackage[colorlinks=true, citecolor=darkBlue, linkcolor=darkBlue, urlcolor=blue]{hyperref}
\usepackage{amssymb}
\usepackage{tabularx}
\usepackage{booktabs}
\definecolor{darkBlue}{rgb}
{0.08, 0.13, 0.4}

\usepackage{cancel}

\usepackage{tikz}
\usetikzlibrary{positioning, arrows.meta}
\usepackage{pgfplots}

\definecolor{THc}{rgb}{0.9,0.3,0.2}

\newcommand{\canc}[1]{}

\begin{document}

\title{Nonequilibrium dynamics in competing quantum-classical quasiperiodic systems}
\author{F. Perciavalle}
\affiliation{Dipartimento di Fisica, Universit\`a della Calabria, 87036 Arcavacata di Rende (CS), Italy}
\affiliation{INFN, Sezione LNF, gruppo collegato di Cosenza}

\author{G. Jacucci}
\affiliation{Dipartimento di Fisica, Universit\`a della Calabria, 87036 Arcavacata di Rende (CS), Italy}
\affiliation{INFN, Sezione LNF, gruppo collegato di Cosenza}
\affiliation{CNR Nanotec, Istituto di Nanotecnologia S.S. Cosenza, Rende (CS)}

\author{F. Plastina}
\affiliation{Dipartimento di Fisica, Universit\`a della Calabria, 87036 Arcavacata di Rende (CS), Italy}
\affiliation{INFN, Sezione LNF, gruppo collegato di Cosenza}

\author{N. Lo Gullo}
\affiliation{Dipartimento di Fisica, Universit\`a della Calabria, 87036 Arcavacata di Rende (CS), Italy}
\affiliation{INFN, Sezione LNF, gruppo collegato di Cosenza}

\date{\today}

\begin{abstract}
A particle subject to competing coherent and dissipative hopping displays active-like transport ``fuelled'' by quantum coherence: its mean-square displacement crosses over from diffusion to a superdiffusive regime and then to diffusion enhanced by a factor set by the Péclet number. However, the mean-square displacement alone cannot reveal whether this activity is genuinely quantum, as it depends only on the populations, which are driven by currents that are in turn fed by coherences at increasing distances. Using the generalized Aubry–André–Harper model, we probe each of these levels across the extended, critical, and localized phases. When only nearest-neighbor coherences are retained, the ratio of coherent to incoherent currents equals the squared Péclet number for any nearest-neighbor Hamiltonian, so deviations signal longer-range coherence. Transport becomes coherence-dominated in every phase, even the localized one. The phases develop distinct density matrices that persist once all of them spread diffusively: the same transport law hides different quantum states. The coherence extent collapses to nearest neighbors in the homogeneous chain but keeps growing in the critical phase. In the localized phase, the dissipation rate decides whether coherence-dominated transport appears at all. A concise account of the main results is given in a companion paper~\cite{perciavalle2026companionShort}.
\end{abstract}

\maketitle

\section{Introduction}
Active matter describes systems whose constituents continuously draw energy
from their environment and convert it into motion~\cite{bechinger2016active,
burgardt2026quantum, shaebani2020computational, ramaswamy2017active,
vrugt2026colloquium, ramaswamy2010mechainics, fodor2016how}.
In recent years, this phenomenology has been extended to the quantum realm,
where active-like behavior arises from mechanisms such as stochastic tuning
of potentials, conversion of heat or dissipation into motion, and engineered
quantum flocks~\cite{penner2025heat, gipouloux2026active, vrugt2026colloquium,
antonov2025engineering, romano2026self, steiner2026active,
antonov2026modeling, khasseh2025active, lee2026anomalous, assent2026quantum, brunner2026quantum, yamagishi2024proposal}.
In this context, Ref.~\cite{gipouloux2026active} showed that when coherent
hopping competes with dissipative hopping, a particle crosses over from
diffusion to a superdiffusive, active-like regime and eventually to enhanced
diffusion, with a P\'eclet number quantifying the enhancement. Quantumness
thus acts as a ``fuel'' for activity~\cite{perciavalle2026companionShort}.

A quantity commonly used to characterize the transport properties of a system is the mean-square displacement (MSD). However, the MSD alone cannot establish whether an observed enhancement is genuinely due to quantum effects. The density-matrix elements obey, in fact, a hierarchy of equations of motion organized by the distance between the sites they connect: the MSD depends only on the populations, which are driven by currents set by the nearest-neighbor coherences, which are in turn fed by coherences at larger distances. Each level of this hierarchy carries information that the levels below it cannot access. Spatial coherence is therefore a precious source of information: it helps classify the phases of the system and provides deeper insight into the possible origin of the enhancement. In particular, at the level of the currents, transport splits exactly into two channels: a coherent one, carried by the nearest-neighbor coherences, and an incoherent one, carried by the stochastic redistribution of the populations.
Their ratio defines the quantum-to-classical current ratio
(QCCR) $\mathcal{R}$, which probes the mechanism behind the spreading
rather than its rate. When the hierarchy is truncated at nearest
neighbors, $\mathcal{R}$ saturates to the squared P\'eclet number,
$\mathcal{R}_*\simeq\mathrm{Pe}^2$: coherent currents are directly
responsible for the enhancement of diffusion. Under appropriate
assumptions, this holds for the homogeneous chain and, as we show, for any
nearest-neighbor hopping Hamiltonian with external inhomogeneous density
field, whereas longer-range coherence generically breaks it, a deviation
captured by the coherence extent $\xi$ of the density matrix~\cite{perciavalle2026companionShort}.

The quantum system, moreover, is not featureless: a Hamiltonian can support
extended, critical, or localized eigenstates~\cite{han1994critical,
wang2021many, roy2025many, anderson1958absence, nandkishore2015many,
luitz2014universal, duthie2022anomalous, chiaracane2021quantum,
guo2024multiple, settino2020emergence}, with very different coherence
structures. Here we provide a complete, time-resolved picture of how this
structure shapes the dissipative dynamics, using the generalized
Aubry--Andr\'e--Harper (GAAH) model~\cite{han1994critical,
li2023observation, wang2021many, roy2025many, ahmed2025supervised,
qi2023multiple, bhakuni2024noise} to tune the Hamiltonian across its three
phases at fixed dissipation. The phases are indistinguishable at short
times, develop markedly different coherence patterns at intermediate times,
and retain them in the enhanced-diffusive regime, even though all of them
recover diffusive spreading. Coherence-dominated transport emerges in all
phases, even deep in the localized one where coherence is confined to a few
sites, while in the critical phase the largest coherence extent comes with
a weaker coherent current dominance than in the extended phase. These features are
free of finite-size effects, and the dissipation rate controls whether
coherence-dominated transport emerges at all in the localized phase.

The paper is organized as follows. In Sec.~\ref{sec:classical} we
introduce the dissipative hopping model, recall its active-like
phenomenology, and derive the hierarchy of equations for the density-matrix
elements. In Sec.~\ref{sec:homogeneous} we
truncate the hierarchy for the homogeneous chain, derive the memory
relation between incoherent and coherent channels, and use its short- and long-time limits
to define the QCCR and relate it to the P\'eclet number. In
Sec.~\ref{sec:pe-qccr} we show that this correspondence extends to
inhomogeneous nearest-neighbor Hamiltonians within the truncated
hierarchy, where the populations map onto a random-barrier walk, and that
it breaks down once longer-range coherences are retained. In
Sec.~\ref{sec:coherence-extent} we introduce the coherence fraction and the
coherence extent, which resolve these higher levels of the hierarchy. In
Sec.~\ref{sec:gaah} we introduce the GAAH model and characterize its phase
diagram through both the fractal dimension and the coherence extent of the
Hamiltonian eigenstates. In Sec.~\ref{sec:dmat} we analyze the real-space
structure of the density matrix in the three phases at times sampling the
onset of coherence buildup, the superdiffusive window, and the
enhanced-diffusive regime. In Sec.~\ref{sec:dynamics} we follow the full
time evolution of the QCCR and of the coherence extent across the phase
diagram, and show that the resulting dynamical regimes are free of
finite-size effects. In Sec.~\ref{sec:diss} we analyze how the dissipation
rate affects transport and coherence in each phase. Finally, in
Sec.~\ref{sec:conclusions} we summarize our results and discuss
experimental perspectives.
 
\section{Activity in dissipative hopping models}
\label{sec:classical}

We consider a model of a single particle moving on a one-dimensional chain,
described by fermionic creation and annihilation operators $\hat c_j^\dagger$
and $\hat c_j$. The single-particle basis is given by the states
$\ket{j}=\hat c_j^\dagger\ket{0}$, corresponding to a particle localized at
site $j$. A general state in this sector can be written as
\begin{equation}
\rho(t)=\sum_{i,j}\rho_{ij}(t)\ket{i}\bra{j},
\end{equation}
where the diagonal elements $\rho_{jj}$ describe the probability of finding
the particle at site $j$, while the off-diagonal elements $\rho_{ij}$ encode
spatial coherences.

The dynamics is governed by a Lindblad master equation comprising both
coherent Hamiltonian evolution and dissipative hopping processes. To isolate
the contribution of dissipation, we first set $\hat H=0$. The master
equation then reduces to
\begin{equation}
\frac{\partial \rho}{\partial t}
=\frac{\Gamma}{2} \left(\sum_j\mathcal{D}[\hat c_j^\dagger\hat c_{j+1}]\rho
+\sum_j\mathcal{D}[\hat c_j\hat c_{j+1}^\dagger]\rho \right),
\label{eq:classical-lindblad}
\end{equation}
where
$
\mathcal{D}[\hat{\mathcal O}]\rho
=\hat{\mathcal O}\rho\hat{\mathcal O}^\dagger
-\frac{1}{2}\{\hat{\mathcal O}^\dagger\hat{\mathcal O},\rho\}
$
is the standard Lindblad dissipator. The two jump operators,
$\hat c_j^\dagger\hat c_{j+1}$ and $\hat c_j\hat c_{j+1}^\dagger$,
describe incoherent hopping from $j+1$ to $j$ and from $j$ to $j+1$,
respectively, each with rate $\Gamma/2$, so that the particle undergoes
symmetric incoherent hopping with total rate $\Gamma$.

For an initially localized particle, $\rho(0)=\ket{j_0}\bra{j_0}$, the
state remains diagonal in the position basis throughout the evolution:
dissipation alone does not generate spatial coherences, and the populations
$n_j(t)=\rho_{jj}(t)$ obey the classical diffusion equation
\begin{equation}
\frac{d n_j}{dt}
=\frac{\Gamma}{2}\left(n_{j-1}+n_{j+1}-2n_j\right),
\end{equation}
with a mean-square displacement (MSD) growing linearly in time,
\begin{equation}
\mathrm{MSD}(t)
=\sum_j (j-j_0)^2 n_j(t)
=\Gamma t.
\label{eq:classical-diffusion}
\end{equation}
Purely dissipative hopping thus provides a classical diffusive baseline.

We now add back the Hamiltonian, so that the dynamics is governed by the
full Lindblad master equation~\cite{fazio2025many, gipouloux2026active,
passarelli2026range}
\begin{equation}
\frac{\partial \rho}{\partial t}
=-i[\hat H,\rho]
+\frac{\Gamma}{2} \left(\sum_j\mathcal{D}[\hat c_j^\dagger\hat c_{j+1}]\rho
+\sum_j\mathcal{D}[\hat c_j\hat c_{j+1}^\dagger]\rho \right).
\label{eq:lindblad}
\end{equation}
The Hamiltonian now generates coherences between neighboring (and, more
generally, arbitrarily separated) sites, which dissipation
tends to erase. The competition between these two processes underlies all
the physics discussed in this paper.

Its most direct signature is visible already in the MSD. As shown in
Ref.~\cite{gipouloux2026active} for the homogeneous chain,
$\hat H=J\sum_j(\hat c_j^\dagger\hat c_{j+1}+\mathrm{H.c.})$, a particle initially localized at $j_0$ crosses over
between three regimes,
\begin{equation}
\mathrm{MSD}(t)\sim
\begin{cases}
\Gamma t, & t\ll t_{\rm sup}\equiv\Gamma/J^2,\\[2pt]
2 D_\alpha t^\alpha, & t_{\rm sup}\ll t\ll t_{\rm enh}\equiv1/\Gamma,\\[2pt]
\Gamma(1+\mathrm{Pe}^2)t, & t\gg t_{\rm enh}.
\end{cases}
\label{eq:msd-regimes}
\end{equation}
At short times the spreading is indistinguishable from the classical
diffusion of Eq.~\eqref{eq:classical-diffusion}. Beyond $t_{\rm sup}$,
coherent hopping drives a superdiffusive regime with $\alpha\lesssim2$,
which persists up to the dissipation timescale $t_{\rm enh}=1/\Gamma$.
Beyond it, spreading becomes diffusive again, but enhanced by a factor
$1+\mathrm{Pe}^2$ with respect to the classical baseline, the P\'eclet
number $\mathrm{Pe}$ quantifying the enhancement.

This diffusive--superdiffusive--diffusive crossover is what motivates the
active-matter interpretation. In active Brownian motion, self-propulsion
adds a directed transport channel on top of ordinary diffusion, with the
P\'eclet number measuring its relative weight~\cite{bechinger2016active};
here, coherent hopping supplies an analogous channel on top of the
incoherent background of Eq.~\eqref{eq:classical-lindblad}. In this sense,
quantum coherence acts as a fuel for active-like motion: a resource absent in
the classical limit, whose buildup and decay control the superdiffusive
window and whose survival at long times controls the enhancement of
diffusion.

Eq.~\eqref{eq:msd-regimes}, however, concerns the populations alone,
and makes no reference to the coherences $\rho_{ij}$, $i\neq j$, that
mediate the spreading. To connect active-like spreading to the underlying
quantum dynamics, we need two further pieces of information: how much of
the transport is carried by coherent versus incoherent processes, which is
set by the nearest-neighbor coherences entering the particle current, and
how coherence is distributed over larger separations, which requires the
full density matrix. Both follow from the equations of motion of
$\rho_{ij}$, which we now derive.

\subsection{Hierarchical equations of coherence}
\label{sec:hierarchy}
The dynamics of coherences is encoded in the time evolution of the density-matrix elements,
whose dynamics is governed by the Hamiltonian. 
A general nearest-neighbor Hamiltonian takes the form:
\begin{equation}
\hat H=\sum_{j=1}^{L-1}J_j\left(\hat c_j^\dagger\hat c_{j+1}+\mathrm{H.c.}\right)
+\sum_{j=1}^{L}h_j\hat n_j ,
\label{eq:H_general}
\end{equation}
which includes both the homogeneous chain ($J_j=J$, $h_j=0$) and the
GAAH of Sec.~\ref{sec:gaah}. In the
single-particle sector, the elements
$\rho_{ij}=\langle\hat c_j^\dagger\hat c_i\rangle$ obey a closed set of
linear equations, Eq.~\eqref{eq:eom_rho}, derived in
Appendix~\ref{app:eom}.

For $i=j$, these equations reduce to the discrete continuity equation
\begin{equation}
\partial_t n_j=-\left(\mathcal I_j-\mathcal I_{j-1}\right),
\qquad
\mathcal I_j=\mathcal I^{\rm coh}_j+\mathcal I^{\rm inc}_j,
\label{eq:continuity_general}
\end{equation}
with the standard decomposition of the particle current into a coherent and
an incoherent part~\cite{gipouloux2026active},
\begin{equation}
\mathcal I^{\rm coh}_j=iJ_j\left(\rho_{j+1,j}-\rho_{j,j+1}\right),
\qquad
\mathcal I^{\rm inc}_j=\frac{\Gamma}{2}\left(n_j-n_{j+1}\right).
\label{eq:currents_general}
\end{equation}
The coherent component is set by nearest-neighbor coherences and does not
depend on the on-site potential; the incoherent one depends on populations
only.

The equations of motion of the off-diagonal elements, $i\neq j$,
highlight the interdependence of the correlations between sites at different distances. Grouping the density-matrix
elements by the distance $r=|i-j|$ between the sites they connect, those at distance $r\geq1$ obey
\begin{align}
\partial_t\rho_{i,i+r}
={}&
-\left[i(h_i-h_{i+r})+\Gamma\right]\rho_{i,i+r}
\nonumber\\
&-i\left(J_i\,\rho_{i+1,i+r}-J_{i+r-1}\,\rho_{i,i+r-1}\right)
\nonumber\\
&-i\left(J_{i-1}\,\rho_{i-1,i+r}-J_{i+r}\,\rho_{i,i+r+1}\right),
\label{eq:hierarchy_explicit}
\end{align}
with $\rho_{i+r,i}=\rho_{i,i+r}^*$. Denoting by
$\rho^{(r)}$ a generic element at distance $r$, this is a hierarchy,
\begin{equation}
\partial_t\rho^{(r)}
=
\mathcal F_r\left[\rho^{(r-1)},\rho^{(r)},\rho^{(r+1)}\right],
\label{eq:hierarchy}
\end{equation}
in which each ingredient of the dynamics plays a distinct role. Dissipation damps all coherences at the same rate $\Gamma$
without coupling different distances; the on-site potential only detunes
them; hopping alone couples distance $r$ to $r\pm1$. The populations
($r=0$) act as the source of the whole hierarchy.

The hierarchy makes explicit why population dynamics is an incomplete
description of quantum activity. The MSD depends on $\rho^{(0)}$ alone; the
populations are driven by the currents, i.e., by $\rho^{(1)}$; the currents
are in turn fed by $\rho^{(2)}$, and so on. The spatial structure of
coherence reaches particle spreading only through successive levels of
Eq.~\eqref{eq:hierarchy}, so that two dynamics with similar populations, and
hence a similar MSD, can differ substantially at the higher levels: in how
much of the transport is carried by coherences, and in how far coherence
extends in space. Characterizing quantum activity therefore requires
resolving each level: the currents, through
Eq.~\eqref{eq:currents_general}, and the spatial structure of the density
matrix beyond nearest neighbors. As we show in the following, a
truncation of the hierarchy at nearest neighbors yields a memory relation
between the two currents (Sec.~\ref{sec:homogeneous}), accurate for the
homogeneous chain at long times; for structured Hamiltonians, instead,
longer-range coherences are generically appreciable and the full hierarchy
must be retained.

\section{Homogeneous chain: memory relation and current ratio}
\label{sec:homogeneous}
\label{sec:qccr}
 
We first consider the homogeneous chain, $J_j=J$ and $h_j=0$, in which the
results of Ref.~\cite{gipouloux2026active} were originally obtained. Here the
hierarchy can be truncated at nearest neighbors, and the competition between
the two transport channels analyzed in closed form; this case will serve as
the reference point for everything that follows.
 
The coherent current, Eq.~\eqref{eq:currents_general}, vanishes on any state
without nearest-neighbor coherence, whereas the incoherent current depends on
populations only. To compare the two channels, we introduce the
quantum-to-classical current ratio (QCCR),
\begin{equation}
\mathcal R(t)=\frac{\mathcal I^{\rm coh}(t)}{\mathcal I^{\rm inc}(t)},
\qquad
\mathcal I^{\rm coh(inc)}(t)=\sum_j\big|\mathcal I_j^{\rm coh(inc)}(t)\big|,
\label{eq:qccr-def}
\end{equation}
such that $\mathcal R(t)\gg1$ ($\ll1$) signals coherence- (incoherence-)
dominated transport. Unlike the MSD, which reports only how fast the particle
spreads, the QCCR probes the mechanism behind the spreading. The absolute
values prevent currents flowing in opposite directions on different bonds
from canceling, so that $\mathcal I^{\rm coh(inc)}$ measures the total
transport carried by each channel.
 
Neglecting $\rho^{(r \geq 2)}$ in Eq.~\eqref{eq:hierarchy_explicit}, the
nearest-neighbor coherence is driven by the incoherent current alone, and
the two currents obey the memory relation (Appendix~\ref{app:memory})
\begin{equation}
\mathcal{I}_j^{\rm coh}(t)
=
\frac{4J^2}{\Gamma}
\int_0^t dt'\,e^{-\Gamma (t-t')}\,
\mathcal{I}_j^{\rm inc}(t').
\label{eq:general_current_relation}
\end{equation}
The coherent current is thus fed by the history of the incoherent one,
through an exponential memory kernel of width $1/\Gamma$. Two limits follow.
At short times, the coherent current builds up linearly while the incoherent
one stays close to its initial value, so that $\mathcal R(t)\sim t$, and the
two become comparable at $t_{\rm cross}\simeq\Gamma/4J^2=t_{\rm sup}/4$ (see Appendix~\ref{app:memory}).
This ordering of timescales has a transparent interpretation in terms of
the quantum fuel. Quantumness first takes over the transport mechanism: at
$t_{\rm cross}$ the current is already carried mainly by coherences, while
the MSD is still indistinguishable from classical diffusion. Quantumness
takes time to manifest in the spreading, and the superdiffusive
scaling emerges only later, at $t_{\rm sup}\simeq4\,t_{\rm cross}$. Activity
is thus the delayed, population-level manifestation of a quantum takeover
that has already occurred at the level of the currents. At long times, $t_*\gg t_{\rm enh}$, the incoherent current varies
slowly on the scale $1/\Gamma$, and the memory kernel can be replaced by its
weight,
\begin{equation}
\mathcal{I}_j^{\rm coh}(t_*)
\simeq
\frac{4J^2}{\Gamma^2}\,
\mathcal{I}_j^{\rm inc}(t_*).
\label{eq:current_ratio}
\end{equation}
The long-time QCCR therefore approaches $\mathcal R_*\simeq4J^2/\Gamma^2$,
which coincides with the squared P\'eclet number extracted from the
MSD~\cite{gipouloux2026active}, $\mathcal R_*\simeq\mathrm{Pe}^2$. The same
ratio then fixes the enhancement of diffusion,
\begin{equation}
D_{\rm enh}=\frac{\Gamma}{2}(1+\mathrm{Pe}^2)\simeq\frac{\Gamma}{2}(1+\mathcal R_*),
\label{eq:Denh}
\end{equation}
with $\mathrm{MSD}\simeq2D_{\rm enh}t$. Coherence is thus not merely
correlated with enhanced diffusion, but responsible for it, acting as a
quantum counterpart of self-propulsion.

\section{P\'eclet number and current ratio for structured Hamiltonians}
\label{sec:pe-qccr}

The relation $\mathcal R_*\simeq\mathrm{Pe}^2$ has so far been established
only for the homogeneous chain. Its origin is that, at long times, in the enhanced-diffusive regime, the
coherent current on each bond is proportional to the incoherent one,
Eq.~\eqref{eq:current_ratio}. We now show that, at long times, this proportionality, and
with it the correspondence, extends to any Hamiltonian of the form of
Eq.~\eqref{eq:H_general} as long as the hierarchy is truncated at nearest
neighbors, and that it is lost once longer-range coherences are retained.

Neglecting $\rho^{(r\geq2)}$ in Eq.~\eqref{eq:hierarchy_explicit}, and
assuming as in the homogeneous case that the incoherent current varies
slowly compared with the relaxation of the coherences, the two currents
remain proportional bond by bond (Appendix~\ref{app:pe-qccr}),
\begin{equation}
\mathcal I^{\rm coh}_j\simeq\kappa_j\,\mathcal I^{\rm inc}_j,
\qquad
\kappa_j=\frac{4J_j^2}{\Gamma^2+\Delta_j^2},
\label{eq:local-kappa}
\end{equation}
with $\Delta_j=h_j-h_{j+1}$. The local conversion factor $\kappa_j$ is
enhanced by hopping and suppressed by the on-site detuning; for $J_j=J$ and
$h_j=0$, it reduces to $4J^2/\Gamma^2$.

The P\'eclet number and the QCCR can now be computed independently, from
populations and from currents, respectively. On the one hand, the total
current becomes $\mathcal I_j\simeq -W_j\,(n_{j+1}-n_j)$, with
$W_j=\frac{\Gamma}{2}(1+\kappa_j)$: the populations perform a classical
random walk with bond-dependent rates, the random-barrier
model~\cite{denteneer1984diffusion}. At long times, this model diffuses
with an effective rate given by the inverse of the mean of
$W_j^{-1}$~\cite{denteneer1984diffusion}, which yields (Appendix~\ref{app:pe-qccr})
\begin{equation}
\mathrm{Pe}^2
=
\Big\langle\!\!\Big\langle\frac{1}{1+\kappa_j}\Big\rangle\!\!\Big\rangle^{-1}-1 ,
\label{eq:pe-harmonic}
\end{equation}
where $\langle\!\langle\cdot\rangle\!\rangle$ denotes the average over
bonds. On the other hand, since $\kappa_j\geq0$, coherent and incoherent
currents flow in the same direction on every bond. Moreover, at long times
assuming that the
total current to vary slowly from bond to bond, the QCCR evaluates to the same expression, and (Appendix~\ref{app:pe-qccr})
\begin{equation}
\mathcal R_*\simeq \mathrm{Pe}^2.
\label{eq:qccr-pe-truncated}
\end{equation} The correspondence
is therefore not a property of the homogeneous chain: within the truncated
hierarchy, it holds for arbitrary bond-dependent hopping and on-site
potential. The structure of the Hamiltonian enters only through the value
of $\mathrm{Pe}^2$.

When coherences at distance $r\geq2$ are retained, the nearest-neighbor
coherence acquires an additional source set by longer-range coherences,
rather than by the local population imbalance, and the coherent current is
no longer proportional to the incoherent one bond by bond. For the
homogeneous chain, this contribution becomes negligible at long times
(as we will show later), and the correspondence still holds; for a
structured Hamiltonian it generically does not, and
$\mathcal R_*\simeq\mathrm{Pe}^2$ breaks down.

A deviation of $\mathcal R_*/\mathrm{Pe}^2$ from unity in the asymptotic
regime therefore signals that coherences beyond nearest neighbors
contribute to transport. This turns the correspondence into a diagnostic
of the higher levels of the hierarchy, and motivates a quantity that
probes them directly, which we introduce next.

\section{Coherence fraction and coherence extent}
\label{sec:coherence-extent}
To resolve how coherence is distributed across the higher levels of the
hierarchy, we define the $r$-distance coherence fraction
\begin{equation}
\mathcal C(r)=\frac{g(r)}{\sum_{r'=1}^{L-1}g(r')},
\quad
g(r)=\frac{1}{2(L-r)}\sum_{|i-j|=r}|\rho_{ij}|,
\label{eq:coh_fraction}
\end{equation}
i.e., the fraction of the total coherence weight carried by pairs of sites
at distance $r$, with $\sum_{r=1}^{L-1}\mathcal C(r)=1$. Its spread over
distances is quantified by the participation ratio, or coherence extent,
\begin{equation}
\xi=\Big[\sum_r\mathcal C(r)^2\Big]^{-1}.
\label{eq:participation-ratio}
\end{equation}
Its meaning becomes transparent by considering the variance of
$\mathcal C(r)$ around its mean value $1/(L-1)$,
$\sigma_{\mathcal C}^2=(L-1)^{-1}\sum_r[\mathcal C(r)-(L-1)^{-1}]^2$. Since
$\sum_r\mathcal C(r)=1$, one has
$\sum_r\mathcal C(r)^2=(L-1)\,\sigma_{\mathcal C}^2+(L-1)^{-1}$, i.e.,
\begin{equation}
\xi=\frac{L-1}{1+\sigma_{\mathcal C}^2\,(L-1)^2}.
\label{eq:xi-variance}
\end{equation}
Small fluctuations of $\mathcal C(r)$ relative to its mean thus correspond
to coherence spread over many distances, and large fluctuations to
coherence concentrated at few of them. In the limiting cases, $\xi=1$ when
all the coherence is carried by a single distance, and $\xi=L-1$ when it is
uniformly spread over all distances ($\sigma_{\mathcal C}=0$), as for a
fully delocalized pure state, $|\rho_{ij}|=1/L$.

A nonzero $\rho_{ij}$ signals that the particle is in a coherent
superposition of sites $i$ and $j$, rather than in a statistical mixture of
the two. The function $g(r)$ measures the typical strength of such
superpositions at distance $r$: the prefactor, the inverse of the number
$2(L-r)$ of pairs at that distance, makes it an average per pair, not
biased by the larger number of pairs at short distances. $\mathcal C(r)$
then tells over which distances the particle is coherently delocalized, and
$\xi$ is the effective number of such distances.

In the language of the hierarchy, $\mathcal C(1)$ is the weight of the
nearest-neighbor coherences that directly feed the currents, while
$\mathcal C(r\geq2)$ is the weight of the higher levels, which reach the
MSD and the QCCR only indirectly. A coherence concentrated at nearest
neighbors, $\mathcal C(1)\simeq1$ and hence $\xi\simeq1$, is consistent with
the truncation of Sec.~\ref{sec:pe-qccr} and with
$\mathcal R_*\simeq\mathrm{Pe}^2$; a substantial weight at larger distances
signals that the higher levels are active, and that the correspondence may
fail.

Finally, because of the normalization, $\mathcal C(r)$ and $\xi$ are
insensitive to the overall amount of coherence, whose nearest-neighbor part
instead enters the currents: they probe how the coherence is organized
in the state rather than how much of it there is. They are directly
measurable, since single-particle density-matrix elements have been
measured in optical lattices~\cite{ardila2018measuring}, and correlations
with local control in Rydberg atom arrays~\cite{bornet2024enhancing}.

\section{The generalized Aubry--Andr\'e--Harper model}
\label{sec:gaah}

\begin{figure}[!t]
\centering
\includegraphics[width=\columnwidth]{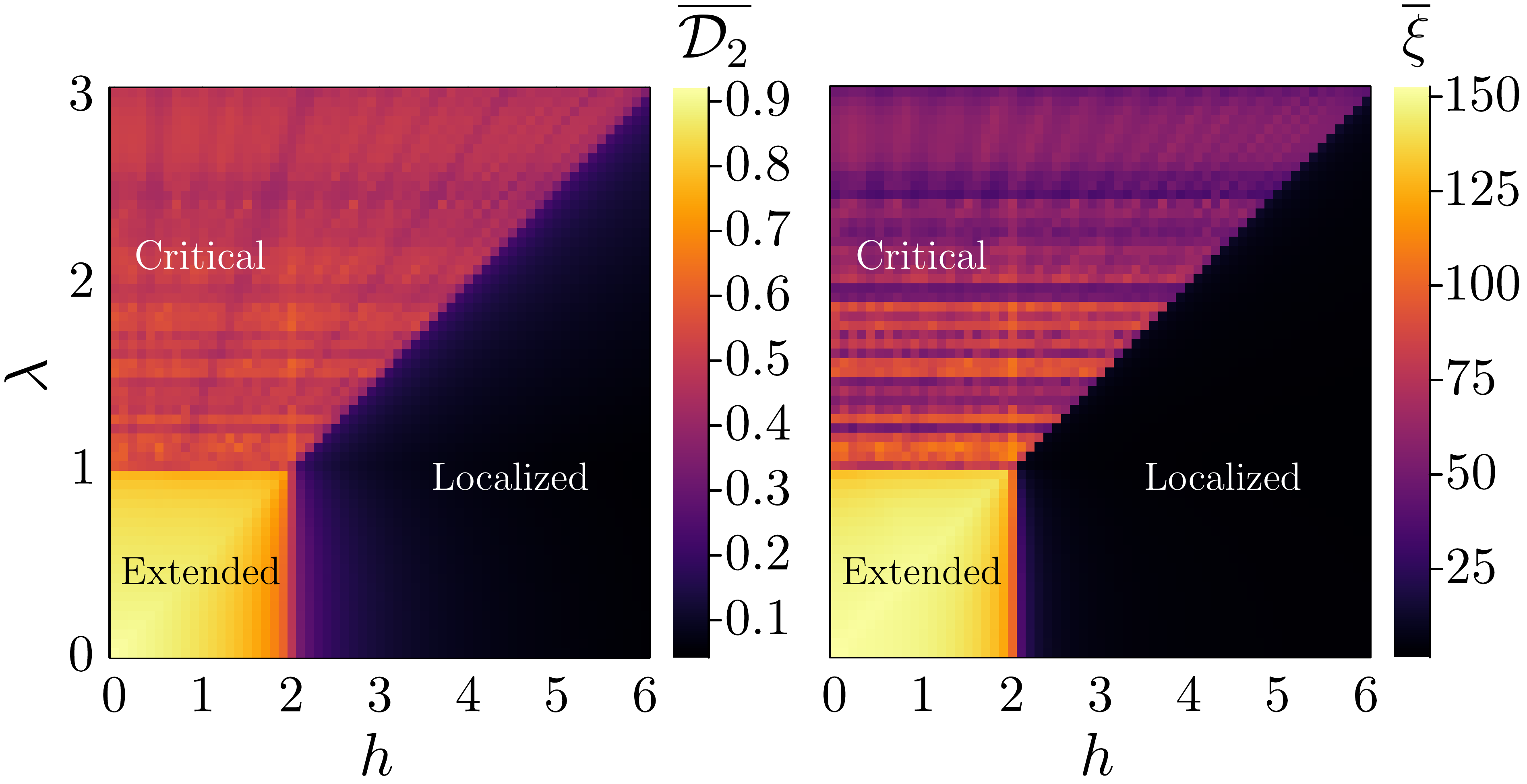}
\caption{Phase diagram of $\hat H$ extracted from the fractal dimension and
the coherence extent, both averaged over the full spectrum. $L=161$,
$\phi=0$.}
\label{fig:diagram}
\end{figure}
\begin{figure*}[!t]
\centering
\includegraphics[width=\textwidth]{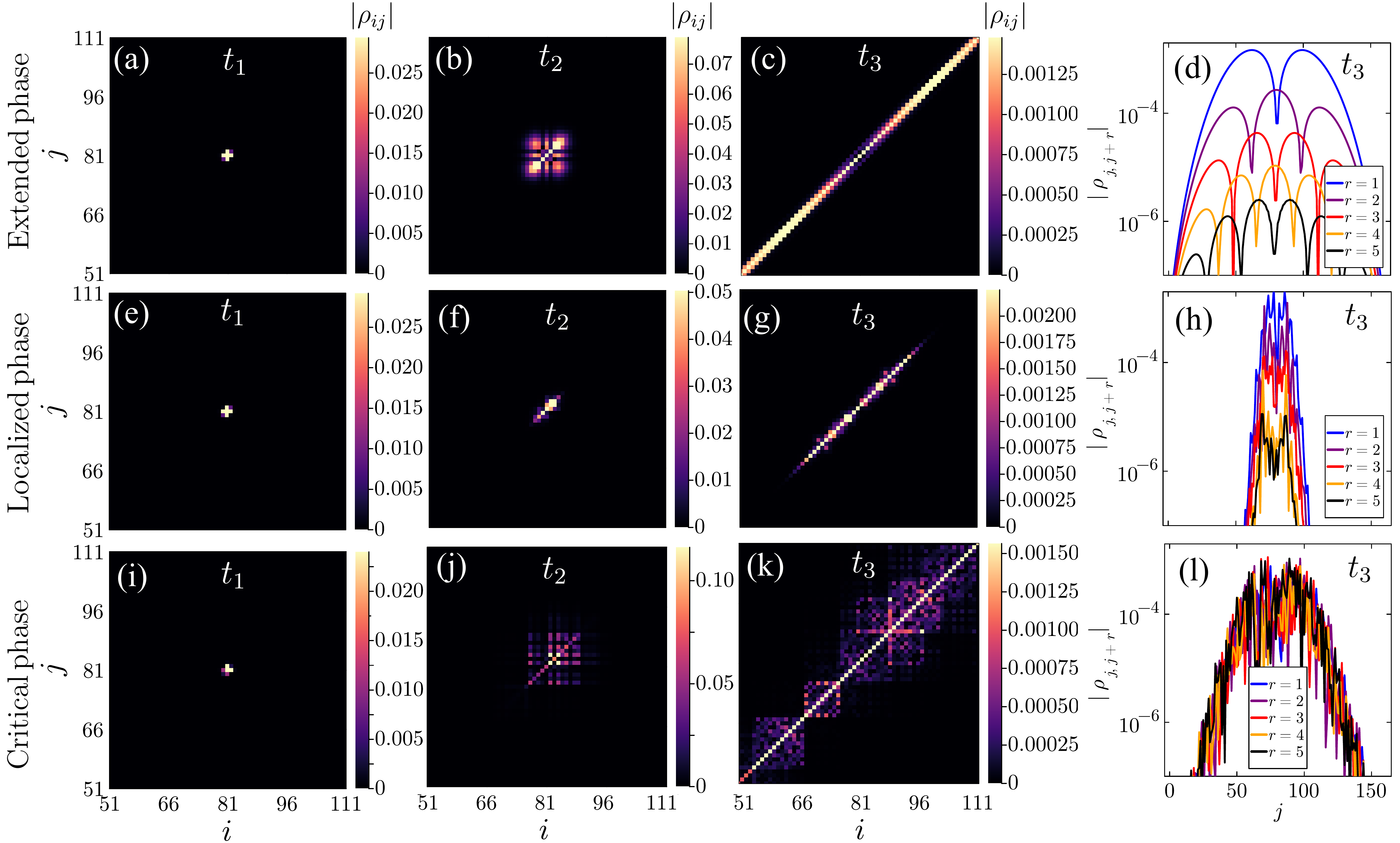} \caption{Density-matrix elements $|\rho_{ij}|$ in the different dynamical regimes and Hamiltonian phases. Panels~(a)-(c) show $|\rho_{ij}|$ at three representative times,
$t_1\simeq t_{\rm cross}/2$, $t_2=t_{\rm enh}/2$
(with $t_{\rm cross}<t_{\rm sup}<t_2<t_{\rm enh}$), and
$t_3=10\,t_{\rm enh}$, respectively.
Panel~(d) shows the spatial dependence of the off-diagonal elements
$|\rho_{j,j+r}|$ for several values of $r$ in the extended phase.
Panels~(e)--(g) show the corresponding density-matrix elements at the same
three times in the localized phase, obtained for $h=5h_c$ and $\lambda=0$,
while panel~(h) shows $|\rho_{j,j+r}|$ for several $r$.
Panels~(i)--(k) present the corresponding results in the quasiperiodically
localized phase, obtained for $h=0$ and $\lambda=5\lambda_c$, for which
$t_1\simeq t_{\rm cross}/6$, and panel~(l) shows $|\rho_{j,j+r}|$ for
several $r$.
The system size is $L=161$, with $\phi=0$ and $\Gamma=1/4$. The colorbar upper limits correspond to the maximum values of the coherences $|\rho_{ij}|$ with $i\neq j$.
}
\label{fig:dmat}
\end{figure*}
To tune the structure of the quantum system in a controlled way, we embed the
same dissipative dynamics in the GAAH
model~\cite{han1994critical, li2023observation, wang2021many, roy2025many,
ahmed2025supervised}, i.e., Eq.~\eqref{eq:H_general} with quasiperiodic
hopping and on-site potential,
\begin{align}
&J_j=\Big|J+\lambda\cos\!\big(2\pi\beta(j+\tfrac12)+\phi\big)\Big|,
\\&
h_j=h\cos(2\pi\beta j+\phi),
\label{eq:gaah}
\end{align}
where $\beta=(\sqrt5-1)/2$ is the inverse golden ratio and $\phi$ is a random
offset phase; we set $J=1$. Already without dissipation, the model hosts
extended, critical (multifractal), and localized phases separated by sharp
transitions~\cite{han1994critical, li2023observation, wang2021many,
roy2025many, ahmed2025supervised, li2022generalized, zhou2026fundamental,
huang2026observation}, whose eigenstates carry structurally different
coherence.

We characterize these phases through the fractal
dimension~\cite{dai2023emergence, fraxanet2022localization} of a normalized
eigenstate $|\psi\rangle=\sum_i\psi_i|i\rangle$,
\begin{equation}
\mathcal{D}_2[\ket{\psi}]
=
-\frac{\ln\!\big(\sum_i|\psi_i|^4\big)}{\ln L},
\end{equation}
with $\mathcal{D}_2\simeq1$, $\mathcal{D}_2\simeq0$, and $0<\mathcal{D}_2<1$
for extended, localized, and critical states, respectively. Its spectral
average, $\overline{\mathcal{D}}_2=L^{-1}\sum_n\mathcal{D}_2[\ket{E_n}]$, with
$\ket{E_n}$ the eigenstates of $\hat H$, yields the phase diagram of
Fig.~\ref{fig:diagram}: an extended--critical transition at $\lambda_c=J$ for
$h=0$, an extended--localized transition at $h_c=2J$ for $\lambda=0$, and a
critical--localized transition along $\lambda=h/2$ for $h>h_c$ and
$\lambda>\lambda_c$.

The same phase structure is captured by the coherence diagnostics of
Sec.~\ref{sec:coherence-extent}. For an eigenstate
$\ket{E_n}=\sum_i\psi_i^n\ket{i}$, the coherences are fixed by the
amplitudes, $|\rho_{ij}|=|\psi_i^n||\psi_j^n|$, and
Eq.~\eqref{eq:participation-ratio} gives its coherence extent $\xi_n$; we
denote by $\overline{\xi}=L^{-1}\sum_n\xi_n$ its spectral average. For a
perfectly extended eigenstate, $|\psi_i^n|=1/\sqrt L$, all distances carry the
same weight and $\xi_n=L-1$, whereas for an exponentially localized one,
$|\rho_{ij}|\propto e^{-|i-j|/\ell}$, the weight concentrates at short
distances and $\xi_n$ approaches unity. As shown in Fig.~\ref{fig:diagram},
the phase diagram obtained from $\overline\xi$ agrees with that from the
fractal dimension, with
$\overline{\xi}^{\rm ext}>\overline{\xi}^{\rm crit}>\overline{\xi}^{\rm loc}$.
This agreement is expected: for pure states, both $\mathcal D_2$ and $\xi$ are
determined by the same amplitudes $|\psi_i^n|$. For the mixed states generated
by the dissipative dynamics, instead, the coherences are no longer fixed by
the populations, and $\xi$ captures information that population-based
quantities such as $\mathcal D_2$ cannot.

\begin{figure*}[!t]
\centering
\includegraphics[width=\textwidth]{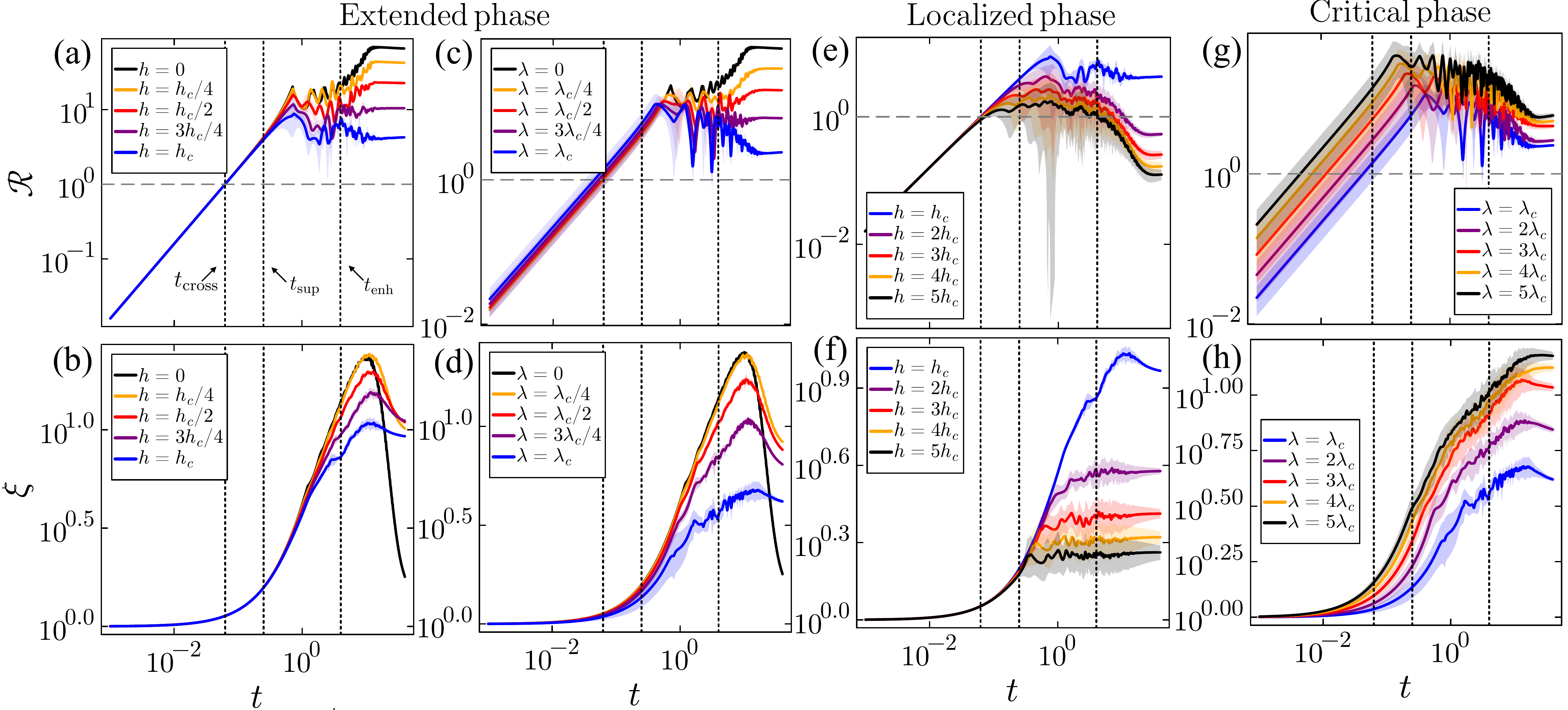} \caption{Dynamics of the QCCR $\mathcal{R}$ and coherence extent $\xi$ across different Hamiltonian phases. Panels (a)–(d) show the extended phase: (a) and (b) depict the dependence on $h$ at $\lambda=0$, while (c) and (d) show the dependence on $\lambda$ at $h=0$. Panels (e)–(f) correspond to the localized phase at $\lambda=0$, whereas panels (g)–(h) correspond to the critical phase at $h=0$. In all panels, $\Gamma=1/4$ and $L=161$. Results are obtained for different $N_\phi=100$ realizations; solid lines denote the averages, and shaded regions indicate the corresponding standard deviations.}
\label{fig:dynamics}
\end{figure*}

\section{Nonequilibrium density matrix structure}
\label{sec:dmat}
Before exploring the QCCR and the coherence extent across the phase
diagram, we look directly at the density matrix in three representative
points: extended ($h=\lambda=0$), localized ($h=5h_c$, $\lambda=0$), and
critical ($h=0$, $\lambda=5\lambda_c$). Figure~\ref{fig:dmat} shows
$|\rho_{ij}|$ at $t_1\simeq t_{\rm cross}/2$, $t_2=t_{\rm enh}/2$, and
$t_3=10\,t_{\rm enh}$, sampling the onset of coherence buildup, the
superdiffusive window, and the enhanced-diffusive regime.

In the critical phase, $\lambda$ enters the hopping amplitudes and
renormalizes the dynamical timescales: the bond-averaged hopping is
$\langle\!\langle J_j\rangle\!\rangle\simeq3.26\,J$ at $\lambda=5\lambda_c$,
roughly three times larger than in the other two cases. We therefore take
the critical-phase snapshot at $t_1\simeq t_{\rm cross}/6$, so as to probe
a comparable stage of the coherence buildup rather than the same absolute
time.

At $t_1$ [Fig.~\ref{fig:dmat}(a),(e),(i)], the particle is still close to
its initial site and the density matrix has essentially the same structure
in all three cases. At the onset of coherence buildup, the
Hamiltonian phase leaves no imprint on the state.

The phases separate sharply by $t_2$ [Fig.~\ref{fig:dmat}(b),(f),(j)]. In
the extended phase, coherence spreads into a symmetric, cross-shaped
pattern centered on $j_0$, extending over several sites both along and
away from the diagonal. In the localized phase, it remains confined to a
narrow region along the diagonal, only marginally wider than at $t_1$, as
expected from exponentially localized eigenstates. In the critical phase,
it develops a fragmented, checkerboard-like structure, extending further
from the diagonal than in the localized case but without the smooth
spreading of the extended one, a first signature of the multifractal
eigenstates.

By $t_3$ [Fig.~\ref{fig:dmat}(c),(g),(k)], all three phases have returned
to diffusive spreading~\cite{perciavalle2026companionShort}, yet their
density matrices remain qualitatively distinct. The restoration of
diffusive scaling does not restore a common quantum state: the same
long-time transport law hides different spatial organizations of
coherence. In the extended phase, $|\rho_{ij}|$ is peaked on the diagonal
with a narrow nearest-neighbor band; in the localized phase, only a thin
band survives immediately around it; in the critical phase, coherence is
organized into extended, block-like clusters along the diagonal, with
sizable weight well beyond nearest neighbors.

Panels (d), (h), (l) make this quantitative through the profiles
$|\rho_{j,j+r}|$ at $t_3$, for $r=1,\dots,5$. In the extended phase
[panel (d)], the curves are well separated on a logarithmic scale, spanning
nearly two decades, with $r=1$ dominant: longer-range coherence is nonzero
but negligible. In the localized phase [panel (h)], the profiles are peaked
around the initial site and decay rapidly with $r$, the dynamical
counterpart of the exponential suppression $|\rho_{ij}|\lesssim e^{-r/\ell}$
of the eigenstate coherences. In the critical phase [panel (l)], instead,
all separations have comparable magnitude across the whole chain, with no
hierarchy in $r$. Only in the critical phase does coherence at longer
range remain comparable to the nearest-neighbor one, which is the origin
of the large coherence extent found in this regime. In the language of
Sec.~\ref{sec:hierarchy}, the extended and localized phases are
effectively truncated at the first levels of the hierarchy at long times,
whereas in the critical phase the higher levels remain active and feed
back into the currents.
\section{Nonequilibrium dynamics of QCCR and coherence extent}
\label{sec:dynamics}
\begin{figure*}[!t]
\centering
\includegraphics[width=\textwidth]{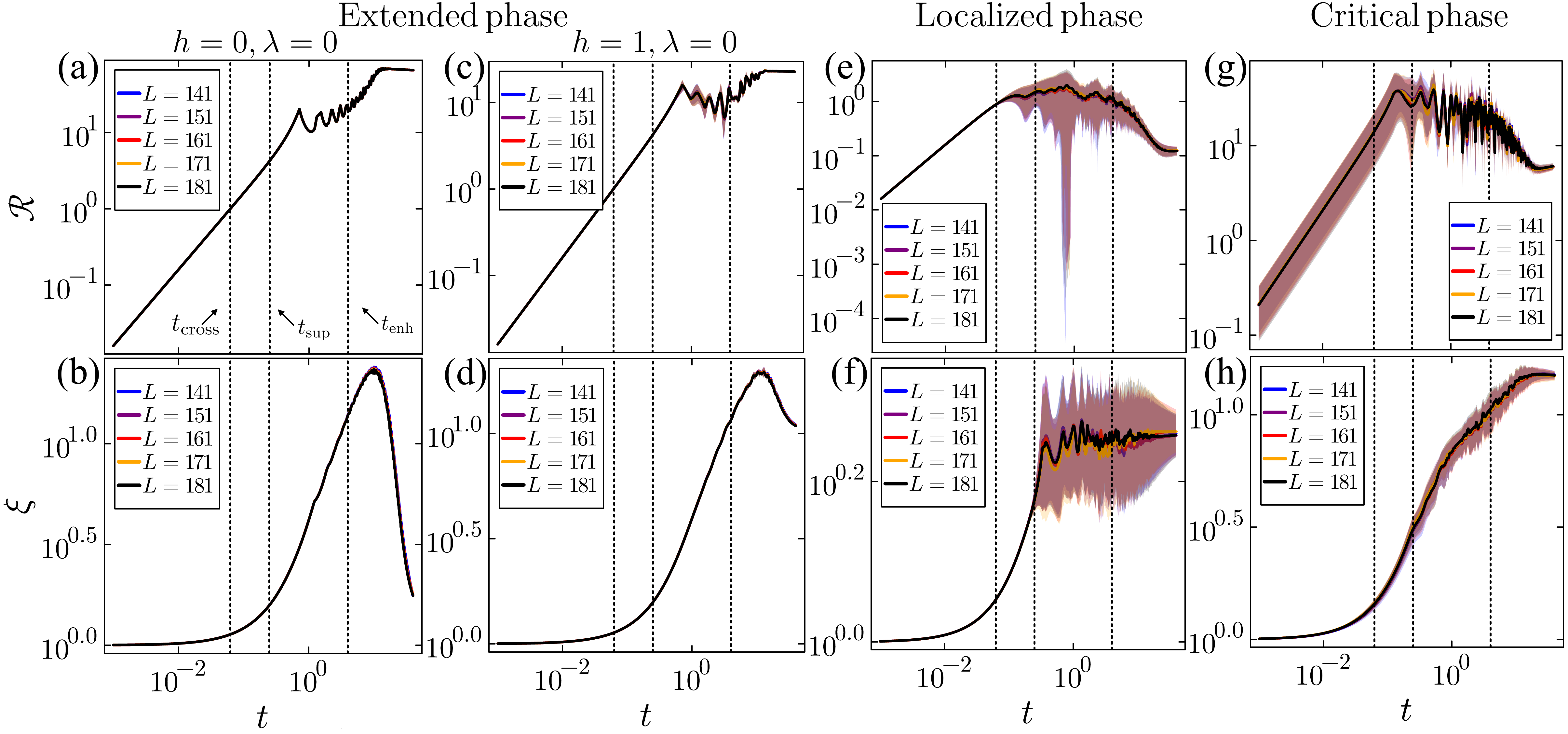} \caption{Dynamics of the QCCR (top panels) and coherence extent (bottom panels) for different system sizes $L$. Panels (a)--(d) correspond to the extended phase, with $h=\lambda=0$ in panels (a) and (b), and $h=1$, $\lambda=0$ in panels (c) and (d). Panels (e) and (f) show the localized phase for $h=10$, $\lambda=0$, while panels (g) and (h) correspond to the critical phase for $\lambda=5$, $h=0$. In all panels, the dissipation strength is fixed at $\Gamma=1/4$, and the dynamics are averaged over $N_\phi=100$ disorder realizations.}
\label{fig:size}
\end{figure*}
We now follow the QCCR $\mathcal R(t)$, which quantifies how much of the
transport is coherent, and the coherence extent $\xi(t)$, which
characterizes how far coherence extends, across the three Hamiltonian
phases (Fig.~\ref{fig:dynamics}). Three main results emerge. First, the
short-time dynamics is the same in all phases. Second, transport
becomes coherence-dominated at intermediate times in every phase,
including deep in the localized one. Third, the relation between
$\mathcal R$ and $\xi$ depends on the phase: they follow the same trend
in the localized and critical phases, but not in the extended one.

\paragraph*{Extended phase.}
In the homogeneous chain [$h=\lambda=0$, black curves in
Fig.~\ref{fig:dynamics}(a)--(d)], $\mathcal R$ first grows linearly,
$\mathcal R(t)\simeq(4J^2/\Gamma)\,t$ [Eq.~\eqref{eq:short_time_coherent}],
crossing unity at $t_{\rm cross}$; its growth then slows down to
$\mathcal R(t)\sim[\ln(t/\tau)]^{\gamma}$, and for $t\gg t_{\rm enh}$ it
saturates at $\mathcal R_*\simeq\mathrm{Pe}_*^2=4J^2/\Gamma^2$
[Eq.~\eqref{eq:current_ratio}]. Meanwhile, $\xi$ grows, peaks after
$t_{\rm enh}$, when $\mathcal R$ saturates, and then drops sharply:
dissipation erases the long-range coherences built during the ballistic
window, and the state relaxes to nearest-neighbor coherence
[cf.~Fig.~\ref{fig:dmat}(d)].

Switching on the quasiperiodic modulation leaves the linear stage of
$\mathcal R$ unaffected, but monotonically suppresses its intermediate
growth and asymptotic value as $h$ increases [Fig.~\ref{fig:dynamics}(a)].
Its effect on the coherence extent is more striking
[Fig.~\ref{fig:dynamics}(b)]. In the homogeneous chain, the rise and
collapse of $\xi$ forms a coherence-extent barrier: during the active
stage, coherence rapidly spreads over many distances, only to be confined
back to nearest neighbors once dissipation takes over. Quasiperiodic
modulation softens the barrier: $\xi$ still reaches a maximum, but its
late-time drop is strongly reduced, so that long-range coherences are
better preserved, though $\xi$ stays far from the delocalized limit
$\xi\simeq L-1$. The ordering is thus inverted: the homogeneous chain,
which builds the highest barrier, is also the one that returns fastest to
nearest-neighbor coherence. The same holds when the hopping is modulated
[Fig.~\ref{fig:dynamics}(c),(d)]. In the extended phase, a large
current ratio can coexist with coherence confined to nearest neighbors,
while better-preserved long-range coherence comes with a smaller current
ratio.

\paragraph*{Localized phase.}
With exponentially localized eigenstates, one would expect transport to
be classical, carried by incoherent hopping alone. The QCCR shows that
this holds only asymptotically [Fig.~\ref{fig:dynamics}(e)]: transport
follows the sequence incoherence-dominated $\to$ coherence-dominated $\to$
incoherence-dominated. The short-time buildup of $\mathcal R$ is the same
as in the extended phase, and even deep in the localized phase
$\mathcal R\gtrsim1$ over an intermediate window. Only later, after strong
realization-dependent fluctuations, does $\mathcal R$ relax to values that
decrease monotonically with $h$ and fall below unity for $h\geq2h_c$.
Meanwhile, $\xi$ shows no barrier [Fig.~\ref{fig:dynamics}(f)]: for
$h>h_c$ it saturates early, at $t\sim t_{\rm sup}$, to a value that
decreases with $h$ [cf.~Fig.~\ref{fig:dmat}(h)]. The transient
coherence-dominated regime thus occurs with coherence confined to a few
sites, since the coherent current only requires nearest-neighbor
coherences. As $h$ increases, $\mathcal R$ and $\xi$ decrease together.
At $h=h_c$, both remain markedly larger, and transport stays
coherence-dominated at all times after $t_{\rm cross}$.

\begin{figure*}[!t]
\centering
\includegraphics[width=\textwidth]{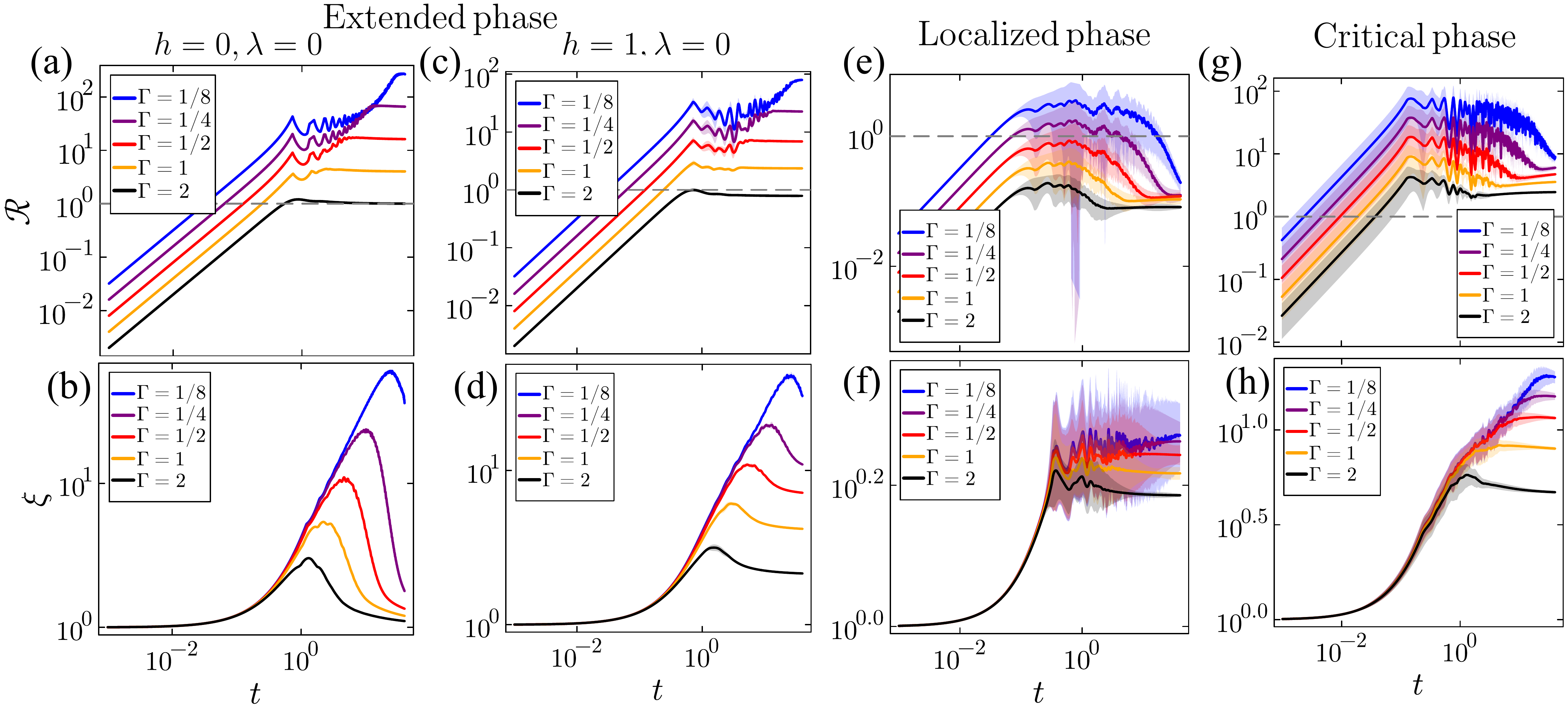} \caption{Dynamics of the QCCR (top panels) and coherence extent (bottom panels) for different values of the dissipation rate $\Gamma$ with $J=1$ fixed. Panels (a)--(d) correspond to the extended phase, with $h=\lambda=0$ in panels (a) and (b), and $h=1$, $\lambda=0$ in panels (c) and (d). Panels (e) and (f) show the localized phase for $h=10$, $\lambda=0$, while panels (g) and (h) correspond to the critical phase for $\lambda=5$, $h=0$. In all panels, $L=161$, and the dynamics are averaged over $N_\phi=100$ disorder realizations.}
\label{fig:gamma}
\end{figure*}
\paragraph*{Critical phase.}
In the critical phase, coherent transport persists at long times but
with a moderate dominance, while coherence extends over the largest
distances among all phases [Fig.~\ref{fig:dynamics}(g),(h)]. Since
$\lambda$ enhances the hopping amplitudes, the short-time QCCR grows with
a larger slope, and the crossover occurs earlier than the homogeneous
$t_{\rm cross}$ (dotted line). After a broad, strongly fluctuating maximum,
$\mathcal R$ decreases for $t\gtrsim t_{\rm enh}$ toward asymptotic values
above, but of the order of, unity. In contrast, $\xi$ keeps growing
throughout the dynamics without presenting a barrier, it increases with $\lambda$, and attains the largest
late-time values among all phases, without reaching a stationary value
within the simulated window. This is the dynamical counterpart of the
block-like structure of Fig.~\ref{fig:dmat}(k). Both $\mathcal R$ and $\xi$
increase with $\lambda$, so that, as in the localized phase, they follow
the same trend.

Overall, the phases are thus distinguished progressively in time. For
$t\lesssim t_{\rm sup}$, all behave alike: $\mathcal R$ grows linearly
while $\xi$ stays essentially flat. For
$t_{\rm sup}\lesssim t\lesssim t_{\rm enh}$, they start to differ: the
growth of $\mathcal R$ slows down in the extended phase and fluctuates
strongly in the localized and critical ones, while $\xi$ keeps growing in
the extended and critical phases and saturates in the localized one.
The sharpest differences emerge only for $t\gtrsim t_{\rm enh}$, in
the enhanced-diffusive regime, precisely where the MSD is diffusive in all
phases: the asymptotic QCCR depends strongly on the phase, the coherence
extent collapses to nearest neighbors only in the homogeneous chain, and
the relation between $\mathcal R$ and $\xi$ becomes phase dependent.

\subsection{Size scaling of QCCR and coherence extent}
\label{subsec:size_scaling}
To check that these regimes and their characteristic times are not
finite-size artifacts, we repeat the protocol for different system sizes
$L$ (Fig.~\ref{fig:size}). In the extended phase, both for $h=0$
[Fig.~\ref{fig:size}(a),(b)] and $h=1$ [Fig.~\ref{fig:size}(c),(d)], the
curves for all sizes collapse onto a single one over the entire time
window: the linear growth of $\mathcal R$, the oscillations after
$t_{\rm sup}$, the late-time plateau, and the position and height of the
maximum of $\xi$ are all size independent. In the localized
[Fig.~\ref{fig:size}(e),(f)] and critical [Fig.~\ref{fig:size}(g),(h)]
phases, realization-to-realization fluctuations are sizeable, but the mean
curves again coincide for all $L$ within the statistical spread.

This is expected, since both the particle packet and the coherence extent
remain much smaller than the system size throughout the simulated window:
coherence is generated and redistributed within a region that never
reaches the chain boundaries. The dynamical regimes and their
characteristic times are therefore intrinsic properties of the model.

\section{Dissipation Tuning and Dynamical Response}
\label{sec:diss}
In Ref.~\cite{gipouloux2026active}, it was shown that the choice $\Gamma=1/4$ is optimal for obtaining active-like dynamics at $h=\lambda=0$. Indeed, increasing $\Gamma$ beyond this value inevitably suppresses activity, leading instead to dissipation-dominated diffusive transport. This naturally raises the question of how $\Gamma$ affects the dynamics across different regions of the phase diagram? To address this question, we analyze, as before, the QCCR and coherence extent, focusing on their dependence on $\Gamma$. Fig.~\ref{fig:gamma} reports $\mathcal{R}(t)$ and $\xi(t)$ for
$\Gamma \in \{1/8, 1/4, 1/2, 1, 2\}$, in the same
representative points of the phase diagram used in Fig.~\ref{fig:size}.

At short times the response to $\Gamma$ is the same in all phases. The
linear buildup $\mathcal{R}(t) \simeq (4J^2/\Gamma)\,t$ shows up as parallel straight lines in log-log
scale, rigidly shifted downward as $\Gamma$ increases. As a result, the
crossover time $t_{\rm cross} \simeq \Gamma/4J^2$ moves to later times for
stronger dissipation. Likewise, the curves of $\xi$ for different $\Gamma$ essentially coincide at early times in
all phases, so the initial growth of the coherence extent is fixed by the
Hamiltonian rather than by dissipation. The phases differ at later times.

\paragraph*{Extended phase.}
In the homogeneous chain [Fig.~\ref{fig:gamma}(a)], the long-time plateau
follows the prediction $\mathcal{R}_* \simeq \mathrm{Pe}^2 = 4J^2/\Gamma^2$. The saturation values scale as $\Gamma^{-2}$,
from $\mathcal{R}_* \simeq 1$ at $\Gamma = 2$ up to $\mathcal{R}_* \simeq 256$
at $\Gamma = 1/8$. The oscillations following $t_{\rm sup}$ become more
pronounced as $\Gamma$ decreases, since the coherent channel is less
damped. At $\Gamma = 2$ the intermediate window (between linear growth and saturation) disappears
altogether, and $\mathcal{R}(t)$ reduces to a simple crossover from a
linear growth to a flat plateau at $\mathcal{R} \simeq 1$, so that the
coherent and incoherent channels carry comparable current at all later
times. The same
picture holds for $h = 1$ [Fig.~\ref{fig:gamma}(c)]. The coherence extent [Fig.~\ref{fig:gamma}(b),(d)] shows the
barrier structure discussed in Sec.~\ref{sec:dynamics}. Both the peak position and height grow as
$\Gamma$ decreases. Stronger dissipation therefore both limits the maximal
coherence extent and speeds up the collapse to nearest-neighbor coherence.

\paragraph*{Localized phase.}
In the localized phase [Fig.~\ref{fig:gamma}(e)], $\Gamma$ decides whether
transport is ever coherence-dominated. For all values of $\Gamma$,
$\mathcal{R}(t)$ displays a hump at intermediate times, before saturation,
signaling that the competition between coherent and incoherent transport
is still ongoing. Only for the weakest dissipation ($\Gamma \lesssim 1/4$)
does the maximum of this hump reach or exceed unity, while for
$\Gamma \geq 1/2$ it remains below it, so that transport stays
incoherence-dominated throughout. At long times $t\gg 1/\Gamma$, the competition is
eventually won by the classical channel. The behavior of $\xi$ [Fig.~\ref{fig:gamma}(f)] does not depend
qualitatively on $\Gamma$. The main difference is that its value in the enhanced-diffusive regime
decreases as $\Gamma$ increases: stronger dissipation further shortens
the distances over which coherence survives.

\paragraph*{Critical phase.}
In the critical phase [Fig.~\ref{fig:gamma}(g)], the renormalized hopping
$\langle \!\langle J_{j} \rangle \!\rangle$ enhances the slope of the initial linear growth of
$\mathcal{R}$, so that the crossover to coherence-dominated transport
occurs earlier for all $\Gamma$. After a broad, strongly fluctuating
maximum, $\mathcal{R}$ relaxes toward values larger than one. Even at
$\Gamma = 2$, $\mathcal{R}$ transiently exceeds unity before relaxing. The coherence extent [Fig.~\ref{fig:gamma}(h)] is largely
sensitive to dissipation. For weak dissipation ($\Gamma \lesssim 1/4$) it
keeps growing throughout the simulated window. For $\Gamma \gtrsim 1$ it
develops a maximum followed by a partial slow decrease, qualitatively like the
extended phase but with a much slower collapse. Strong dissipation thus partly suppresses the broadly
distributed coherence.

\section{Conclusions}
\label{sec:conclusions}
We have studied a single particle subject to competing coherent and
dissipative hopping, a minimal setting in which quantumness acts as the
fuel of active-like transport~\cite{perciavalle2026companionShort}. The
particle spreading cannot reveal whether this activity is quantum: the
MSD depends only on the populations, which sit at the bottom of an exact
hierarchy in which populations are driven by currents and currents by
coherences. We have probed this hierarchy level by level, using the
GAAH model to tune the Hamiltonian across
its extended, critical, and localized phases.

At the level of currents, an exact decomposition into coherent and
incoherent channels defines the quantum-to-classical current ratio. When
the hierarchy is truncated at nearest neighbors, its long-time value
equals the squared P\'eclet number for any nearest-neighbor Hamiltonian:
coherence is not merely correlated with the enhancement of diffusion but
responsible for it, and deviations from this correspondence signal
longer-range coherences. Across the phase diagram, transport becomes
coherence-dominated in every phase, only transiently in the localized
one, even when coherence is confined to a few sites.

At the level of coherences, the phases differ most. Indistinguishable at
short times, they develop distinct density matrices at intermediate times,
and these differences persist in the enhanced-diffusive regime, where all
phases spread diffusively: the same transport law hides different quantum
states. The coherence extent makes this quantitative. In the homogeneous
chain, it rises and then collapses to nearest neighbors, a barrier that
quasiperiodic modulation breaks. In the localized phase, it saturates
early at a few sites; in the critical phase, it keeps growing and reaches
the largest values. Current ratio and coherence extent thus carry
different information: the homogeneous chain has the most coherent
transport, yet ends with the least extended coherence.

These features are free of finite-size effects, since coherence is
generated and redistributed within a region much smaller than the chain.
The dissipation rate, instead, strongly shapes them. In the extended
phase, it sets the height and position of the coherence-extent barrier;
in the critical phase, strong dissipation makes a barrier emerge and
suppresses long-range coherence; in the localized phase, it decides
whether a coherence-dominated regime appears at all.

These predictions are within reach of current experiments. The GAAH model
has been realized with ultracold atoms and superconducting
circuits~\cite{li2023observation,huang2026observation}, and dissipative
hopping can be engineered in cold-atom and Rydberg
platforms~\cite{haga2021liouvillian,schempp2015correlated}. All our
diagnostics require only site populations and two-point correlators,
which have been measured in optical lattices~\cite{ardila2018measuring}
and Rydberg tweezer arrays~\cite{bornet2024enhancing,
perciavalle2023controlled, chen2023continuous}. Several directions remain
open. Long-range hopping, natural in Rydberg platforms, would allow one to
study how coherence built over large distances competes with
quasiperiodic modulation and dissipation~\cite{defenu2023long,
passarelli2026range}. Extending the analysis to interacting systems would clarify whether this
richness beyond populations persists in collective quantum active
dynamics, and how interactions shape key indicators such as the current
ratio and the coherence extent.

\section*{Acknowledgements}
We thank G. Fava, G. Tucci, O. La Rocca, A. Palamara for fruitful discussions. Numerical simulations were performed using the Julia package
\texttt{QuantumToolbox.jl}~\cite{QuantumToolbox.jl2025}. The authors used Claude (Anthropic) and ChatGPT (OpenAI) for assistance with coding and editing. All outputs were checked by the author, who takes full responsibility for the content. We acknowledge support from PNRR MUR Project No. PE0000023-NQSTI through the secondary project ThAnQ.

\appendix

\section{Equations of motion}
\label{app:eom}
 
We derive the equations of motion of the density-matrix elements for the
Hamiltonian of Eq.~\eqref{eq:H_general} and the dissipator of
Eq.~\eqref{eq:lindblad}. In the Heisenberg picture, the one-body operator
$\hat c_i^\dagger\hat c_j$ evolves as
\begin{align}
\partial_t\hat c_i^\dagger\hat c_j
=
i[\hat H,\hat c_i^\dagger\hat c_j]
+\frac{\Gamma}{2}\sum_l\Big(&\mathcal D^\dagger[\hat c_l^\dagger\hat c_{l+1}](\hat c_i^\dagger\hat c_j)
\nonumber\\
+\,&\mathcal D^\dagger[\hat c_l\hat c_{l+1}^\dagger](\hat c_i^\dagger\hat c_j)\Big),
\label{eq:heisenberg}
\end{align}
where $\mathcal D^\dagger[\hat{\mathcal O}](\hat A)\equiv\hat{\mathcal
O}^\dagger\hat A\hat{\mathcal O}-\frac12\{\hat{\mathcal O}^\dagger\hat{\mathcal
O},\hat A\}$ is the adjoint dissipator.
 
\paragraph*{Hamiltonian contribution.}
Using the bilinear commutator identity
$[\hat c_a^\dagger\hat c_b,\hat c_i^\dagger\hat c_j]
=\delta_{bi}\hat c_a^\dagger\hat c_j-\delta_{aj}\hat c_i^\dagger\hat c_b$,
\begin{align}
i[\hat H,\hat c_i^\dagger\hat c_j]
={}&
iJ_{i-1}\hat c_{i-1}^\dagger\hat c_j
+iJ_i\hat c_{i+1}^\dagger\hat c_j
\nonumber\\
&-iJ_{j-1}\hat c_i^\dagger\hat c_{j-1}
-iJ_j\hat c_i^\dagger\hat c_{j+1}
\nonumber\\
&+i(h_i-h_j)\,\hat c_i^\dagger\hat c_j .
\end{align}
 
\paragraph*{Dissipative contribution.}
In the single-particle sector, $\hat c_i^\dagger\hat c_j=|i\rangle\langle j|$. The jump operators then read
$\hat c_l^\dagger\hat c_{l+1}=|l\rangle\langle l+1|$ and
$\hat c_l\hat c_{l+1}^\dagger=-|l+1\rangle\langle l|$, i.e., each bond
carries one jump in each direction; since $\mathcal D^\dagger$ is quadratic
in the jump operator, the sign is irrelevant. Both cases are covered by a
generic jump $\hat L=|a\rangle\langle b|$, which moves the particle from site
$b$ to a neighboring site $a$. For $\hat A=|i\rangle\langle j|$,
\begin{equation}
\hat L^\dagger\hat A\hat L
=
\delta_{ai}\delta_{aj}\,\hat n_b,
\qquad
\left\{\hat L^\dagger\hat L,\hat A\right\}
=
(\delta_{bi}+\delta_{bj})\,\hat c_i^\dagger\hat c_j ,
\end{equation}
where we used that $\hat L^\dagger\hat L=|b\rangle\langle b|$ is a
projector. Summing over all bonds and both directions, the first term gives
$\delta_{ij}(\hat n_{i-1}+\hat n_{i+1})$, while, since in the bulk every site
is the starting point of two jumps, the second gives
$4\,\hat c_i^\dagger\hat c_j$. Hence, in the bulk,
\begin{equation}
\frac{\Gamma}{2}\sum_{\hat L}\mathcal D^\dagger[\hat L]\left(\hat c_i^\dagger\hat c_j\right)
=
\frac{\Gamma}{2}\,\delta_{ij}\left(\hat n_{i-1}+\hat n_{i+1}\right)
-\Gamma\,\hat c_i^\dagger\hat c_j,
\label{eq:Ddag}
\end{equation}
where the sum over $\hat{L}$ accounts for the sums over $l$ of the operators $\hat c_l^\dagger\hat c_{l+1}$ and
$\hat c_l\hat c_{l+1}^\dagger$.
 
\paragraph*{Complete equations.}
Combining both contributions,
\begin{align}
\partial_t\left(\hat c_i^\dagger\hat c_j\right)
={}&
iJ_{i-1}\hat c_{i-1}^\dagger\hat c_j
+iJ_i\hat c_{i+1}^\dagger\hat c_j
\nonumber\\
&-iJ_{j-1}\hat c_i^\dagger\hat c_{j-1}
-iJ_j\hat c_i^\dagger\hat c_{j+1}
\nonumber\\
&+\left[i(h_i-h_j)-\Gamma\right]\hat c_i^\dagger\hat c_j
\nonumber\\
&+\frac{\Gamma}{2}\,\delta_{ij}\left(\hat n_{i-1}+\hat n_{i+1}\right),
\label{eq:eom_operator}
\end{align}
with $J_0=J_L=0$ for open boundary conditions. Taking expectation values,
the density-matrix elements $\rho_{ij}=\langle\hat c_j^\dagger\hat c_i\rangle$
obey
\begin{align}
\partial_t\rho_{ij}
={}&
-i\left(J_i\,\rho_{i+1,j}+J_{i-1}\,\rho_{i-1,j}\right)
\nonumber\\
&+i\left(J_j\,\rho_{i,j+1}+J_{j-1}\,\rho_{i,j-1}\right)
\nonumber\\
&-\left[i(h_i-h_j)+\Gamma\right]\rho_{ij}
\nonumber\\
&+\frac{\Gamma}{2}\,\delta_{ij}\left(\rho_{i-1,i-1}+\rho_{i+1,i+1}\right).
\label{eq:eom_rho}
\end{align}
For $i=j$, Eq.~\eqref{eq:eom_rho} yields the continuity equation,
Eq.~\eqref{eq:continuity_general}, with the currents of
Eq.~\eqref{eq:currents_general}; setting $j=i+r$ yields
Eq.~\eqref{eq:hierarchy_explicit}:
\begin{align}
\partial_t\rho_{i,i+r}
={}&
-\left[i(h_i-h_{i+r})+\Gamma\right]\rho_{i,i+r}
\nonumber\\
&-i\left(J_i\,\rho_{i+1,i+r}-J_{i+r-1}\,\rho_{i,i+r-1}\right)
\nonumber\\
&-i\left(J_{i-1}\,\rho_{i-1,i+r}-J_{i+r}\,\rho_{i,i+r+1}\right).
\end{align}

\section{Memory relation and its limits}
\label{app:memory}
 
For $J_j=J$ and $h_j=0$, neglecting $\rho^{(2)}$ in
Eq.~\eqref{eq:hierarchy_explicit} at $r=1$ gives~\cite{gipouloux2026active}
\begin{equation}
\partial_t\rho_{j,j+1}
=
-\Gamma\,\rho_{j,j+1}
+\frac{2iJ}{\Gamma}\,\mathcal I_j^{\rm inc}.
\label{eq:coherence-eom-truncated}
\end{equation}
For an initially incoherent state, $\rho_{j,j+1}(0)=0$, its solution
is
\begin{equation}
\rho_{j,j+1}(t)
=
\frac{2iJ}{\Gamma}\int_0^t dt'\,e^{-\Gamma(t-t')}\,
\mathcal I_j^{\rm inc}(t'),
\label{eq:coherence-eom-solution}
\end{equation}
and inserting it into
$\mathcal I_j^{\rm coh}=iJ(\rho_{j+1,j}-\rho_{j,j+1})$ yields the memory
relation, Eq.~\eqref{eq:general_current_relation}:
\begin{equation}
\mathcal{I}_j^{\rm coh}(t)
=
\frac{4J^2}{\Gamma}
\int_0^t dt'\,e^{-\Gamma (t-t')}\,
\mathcal{I}_j^{\rm inc}(t').
\end{equation}
 
\paragraph*{Short times.}
Applying the triangle inequality to Eq.~\eqref{eq:general_current_relation},
\begin{equation}
\mathcal{I}^{\rm coh}(t)
\leq
\frac{4J^2}{\Gamma}
\int_0^t dt'\,e^{-\Gamma (t-t')}\,
\mathcal{I}^{\rm inc}(t').
\label{eq:triangle-bound-main}
\end{equation}
For $\Gamma t\ll1$, the incoherent current has not yet relaxed,
$\mathcal{I}^{\rm inc}(t)\simeq\mathcal{I}^{\rm inc}(0)$, so that
\begin{equation}
\mathcal{I}^{\rm coh}(t)
\lesssim
\frac{4J^2}{\Gamma}\,t\,\mathcal{I}^{\rm inc}(0),
\label{eq:short_time_coherent}
\end{equation}
and the two currents become comparable at
\begin{equation}
t_{\rm cross}\sim\frac{\Gamma}{4J^2}=\frac{t_{\rm sup}}{4}.
\label{eq:current_crossover}
\end{equation}
 
\paragraph*{Long times.}
We write Eq.~\eqref{eq:coherence-eom-solution} as
\begin{equation}
\rho_{j,j+1}(t)=\frac{2iJ}{\Gamma}\,e^{-\Gamma t}K_j(t),
\,\,\,
K_j(t)\equiv\int_0^t dt'\,e^{\Gamma t'}\,\mathcal I_j^{\rm inc}(t'),
\label{eq:K_definition}
\end{equation}
and integrate $K_j(t)$ by parts,
\begin{equation}
K_j(t)
=
\frac{e^{\Gamma t}}{\Gamma}\mathcal{I}_j^{\rm inc}(t)
-\frac{\mathcal{I}_j^{\rm inc}(0)}{\Gamma}
-\frac{1}{\Gamma}\int_0^t dt'\,e^{\Gamma t'}\dot{\mathcal{I}}_j^{\rm inc}(t').
\end{equation}
For $\Gamma t\gg1$, the boundary term is exponentially subleading. The
remaining integral is dominated, through the weight $e^{\Gamma t'}$, by times
$t'\sim t$; if $\dot{\mathcal{I}}_j^{\rm inc}$ varies slowly over this range,
it gives $\dot{\mathcal{I}}_j^{\rm inc}(t)\,e^{\Gamma t}/\Gamma$, which
carries the same exponential factor as the leading term and remains
subleading if and only if
\begin{equation}
\big|\dot{\mathcal{I}}_j^{\rm inc}(t)\big|
\ll
\Gamma\big|\mathcal{I}_j^{\rm inc}(t)\big|.
\label{eq:slowly_varying_condition}
\end{equation}
Under this condition,
$K_j(t)\simeq(e^{\Gamma t}/\Gamma)\,\mathcal{I}_j^{\rm inc}(t)$, hence
$\rho_{j,j+1}(t)\simeq(2iJ/\Gamma^2)\,\mathcal{I}_j^{\rm inc}(t)$, and
substituting into the coherent current gives
Eq.~\eqref{eq:current_ratio}:
\begin{equation}
\mathcal{I}_j^{\rm coh}(t_*)
\simeq
\frac{4J^2}{\Gamma^2}\,
\mathcal{I}_j^{\rm inc}(t_*).
\end{equation}

\section{P\'eclet number and current ratio in the truncated hierarchy}
\label{app:pe-qccr}
 
\paragraph*{Local proportionality of the currents.}
Neglecting $\rho^{(r\geq2)}$ in Eq.~\eqref{eq:hierarchy_explicit}, at long times, the
nearest-neighbor coherence $\rho_j\equiv\rho_{j,j+1}$ obeys
\begin{equation}
\partial_t\rho_j
=
-(\Gamma+i\Delta_j)\,\rho_j
+\frac{2iJ_j}{\Gamma}\,\mathcal I^{\rm inc}_j,
\qquad
\Delta_j=h_j-h_{j+1}.
\label{eq:rho1-truncated}
\end{equation}
Proceeding as in Appendix~\ref{app:memory}, with $\Gamma\to\Gamma+i\Delta_j$,
and assuming that the incoherent current varies slowly compared with the
relaxation rate of the coherence,
$|\dot{\mathcal I}^{\rm inc}_j|\ll|\Gamma+i\Delta_j|\,|\mathcal I^{\rm inc}_j|$,
one finds
$\rho_j\simeq(2iJ_j/\Gamma)\,\mathcal I^{\rm inc}_j/(\Gamma+i\Delta_j)$.
Since $\mathcal I^{\rm coh}_j=iJ_j(\rho_{j+1,j}-\rho_{j,j+1})
=2J_j\,\mathrm{Im}\,\rho_j$, this gives Eq.~\eqref{eq:local-kappa}:
\begin{equation}
\mathcal I^{\rm coh}_j\simeq\kappa_j\,\mathcal I^{\rm inc}_j,
\qquad
\kappa_j=\frac{4J_j^2}{\Gamma^2+\Delta_j^2}.
\end{equation}
 
\paragraph*{P\'eclet number.}
Substituting Eq.~\eqref{eq:local-kappa} into the total current,
$\mathcal I_j=\mathcal I^{\rm coh}_j+\mathcal I^{\rm inc}_j$,
\begin{equation}
\mathcal I_j=-W_j\,(n_{j+1}-n_j),
\qquad
W_j=\frac{\Gamma}{2}\left(1+\kappa_j\right),
\label{eq:fick}
\end{equation}
and the continuity equation takes the form
\begin{equation}
\partial_t n_j
=W_j\,(n_{j+1}-n_j)-W_{j-1}\,(n_j-n_{j-1}).
\label{eq:diffusion-local}
\end{equation}
For a uniform rate, $W_j=W$, this gives $\mathrm{MSD}=2Wt$. For
bond-dependent rates, Eq.~\eqref{eq:diffusion-local} is the random-barrier
model, which at long times predicts $\mathrm{MSD}\simeq2D_{\rm enh}\,t$ with
an effective diffusion constant given by the mean of the inverse
rates~\cite{denteneer1984diffusion},
\begin{equation}
D_{\rm enh}
=
\langle\!\langle W_j^{-1}\rangle\!\rangle^{-1}
=
\frac{\Gamma}{2}
\Big\langle\!\!\Big\langle\frac{1}{1+\kappa_j}\Big\rangle\!\!\Big\rangle^{-1},
\label{eq:Denh-rbm}
\end{equation}
where
$\langle\!\langle f_j\rangle\!\rangle\equiv\frac1N\sum_{j=1}^{N}f_j$
denotes the average over bonds, not to be confused with quantum expectation
values or spectral averages. Comparing with the enhanced-diffusive form,
$\mathrm{MSD}\simeq\Gamma(1+\mathrm{Pe}^2)\,t$, gives
Eq.~\eqref{eq:pe-harmonic}:
\begin{equation}
\mathrm{Pe}^2
=
\Big\langle\!\!\Big\langle\frac{1}{1+\kappa_j}\Big\rangle\!\!\Big\rangle^{-1}-1 ,
\end{equation}
 
\paragraph*{Current ratio.}
From Eq.~\eqref{eq:local-kappa},
$\mathcal I_j=(1+\kappa_j)\,\mathcal I^{\rm inc}_j$, so that
\begin{equation}
\mathcal I^{\rm inc}_j=\frac{\mathcal I_j}{1+\kappa_j},
\qquad
\mathcal I^{\rm coh}_j=\frac{\kappa_j}{1+\kappa_j}\,\mathcal I_j .
\end{equation}
Because $\kappa_j\geq0$, the three currents have the same sign on every
bond, and Eq.~\eqref{eq:qccr-def} becomes
\begin{equation}
\mathcal R
=
\frac{\sum_j\frac{\kappa_j}{1+\kappa_j}\,|\mathcal I_j|}
{\sum_j\frac{1}{1+\kappa_j}\,|\mathcal I_j|}.
\label{eq:qccr-weighted}
\end{equation}
For the homogeneous chain, $\kappa_j=\kappa$ and $\mathcal R=\kappa$ exactly.
In general, assuming that the total current varies slowly from bond to bond,
it can be factored out of the sums, and, using
$\kappa_j/(1+\kappa_j)=1-1/(1+\kappa_j)$,
\begin{equation}
\mathcal R_*
=
\frac{1-\langle\!\langle1/(1+\kappa_j)\rangle\!\rangle}
{\langle\!\langle1/(1+\kappa_j)\rangle\!\rangle}
=
\Big\langle\!\!\Big\langle\frac{1}{1+\kappa_j}\Big\rangle\!\!\Big\rangle^{-1}-1 ,
\end{equation}
which coincides with Eq.~\eqref{eq:pe-harmonic}.

\end{document}